\documentclass[aps,prl,reprint,groupedaddress]{revtex4-1}

\usepackage[dvipdfmx]{graphicx}
\usepackage{color}
\newcommand{\ayb}{$\alpha$-YbAlB$_4\,$}
\newcommand{\byb}{$\beta$-YbAlB$_4\,$}
\newcommand{\abyb}{$\alpha$- and $\beta$-YbAlB$_4\,$}

\begin{document}


\title{Effect of Anisotropic Hybridization in YbAlB$_4$ Probed by Linear Dichroism in Core-Level Hard X-ray Photoemission Spectroscopy}


\author{Kentaro Kuga,$^1$ Yuina Kanai,$^{1,2}$ Hidenori Fujiwara,$^{1,2}$ Kohei Yamagami,$^{1,2}$ Satoru Hamamoto,$^{1,2}$ Yuichi Aoyama,$^{1,2}$ Akira Sekiyama,$^{1,2}$ Atsushi Higashiya,$^{1,3}$ Toshiharu Kadono,$^{1,4}$ Shin Imada,$^{1,4}$ Atsushi Yamasaki,$^{1,5}$ Arata Tanaka,$^6$ Kenji Tamasaku,$^1$ Makina Yabashi,$^1$ Tetsuya Ishikawa,$^1$ Satoru Nakatsuji,$^{7,8}$ and Takayuki Kiss$^{1,2}$}
\affiliation{$^1$RIKEN SPring-8 Center, Sayo, Hyogo 679-5148, Japan.\\
$^2$Graduate School of Engineering Science, Osaka University, Toyonaka, Osaka 560-8531, Japan.\\
$^3$Faculty of Science and Engineering, Setsunan University, Neyagawa, Osaka 572-8508, Japan.\\
$^4$College of Science and Engineering, Ritsumeikan University, Kusatsu, Shiga 525-8577, Japan.\\
$^5$Faculty of Science and Engineering, Konan University, Kobe, Hyogo 658-8501, Japan.\\
$^6$Department of Quantum Matter, ADSM, Hiroshima Universitty, Higashi-hiroshima, Hiroshima 739-8530, Japan\\
$^7$The Institute for Solid State Physics, The University of Tokyo, Kashiwa, Chiba 277-8581, Japan.\\
$^8$CREST, Japan Science and Technology Agency (JST), 4-1-8 Honcho Kawaguchi, Saitama 332-0012, Japan}


\date{\today}

\begin{abstract}
We have probed the crystalline electric-field ground states of pure $|J = 7/2, J_z = \pm 5/2\rangle$ as well as the anisotropic  $c$-$f$ hybridization in both valence fluctuating systems \abyb by linear polarization dependence of angle-resolved core level photoemission spectroscopy.
Interestingly, the small but distinct difference between \abyb was found in the polar angle dependence of linear dichroism, indicating the difference in the anisotropy of $c$-$f$ hybridization which may be essential to a heavy Fermi liquid state in \ayb and a quantum critical state in \byb.
\end{abstract}

\pacs{xxx}

\maketitle


Crystalline electric field (CEF) ground state (GS) regulates physical properties at low temperatures and sometimes leads to variety of nontrivial quantum states such as quantum criticality in quadrupolar Kondo lattice system \cite{Cox87,Cox94,Kusunose96}, colossal magnetoresistance and high temperature superconductivity in transition metal oxide \cite{Tokura2000}.
For example in high $T_{\rm{c}}$ cuprate, CEF splitting of Cu 3$d$ electron determines the half filling in $d_{x^2-y^2}$ orbit and the orbit anisotropically hybridizes with neighboring O 2$p$ electrons, mediating electric and magnetic interaction \cite{ZhangRice}.
Such an anisotropic hybridization of the orbit selected by CEF plays essential role of the physical properties in many of transition metal oxides \cite{Tokura2000}.
Accurate estimation of CEF GS and hybridization is indispensable to understand the underlying mechanisms.

In rare earth metal systems, CEF GS and hybridization is also important.
One of the most interesting phenomena is a quantum criticality in \byb which stems from the pure $|J = 7/2, J_z = \pm 5/2\rangle$ CEF GS and the nodal hybridization in momentum space as predicted by A. Ramires, $et\ al.$ \cite{ramires2012beta}.
In this theory, a new type of topological phase transition in Fermi surface is predicted to arise and induces a topological non-trivial vortex metal which shows magnetic properties of $T^{-1/2}$ divergence and $T/B$ scaling.

\byb is a fascinating and mysterious material which is the firstly discovered superconductor in Yb-based heavy fermion systems \cite{nakatsuji08,KugaPRL} and shows quantum criticality without tuning \cite{Matsumoto2011Science} in the strongly valence fluctuating state \cite{ybal-valency}.
As a function of pressure, \byb forms non-Fermi-liquid phase including the quantum criticality at ambient pressure \cite{Tomita2015}.
To understand the underlying mechanism, laser ARPES probed anisotropic hybridization as a momentum dependent Kondo hybridization band which is consistent with nodal hybridization model \cite{laserARPES_YbAlB4}.
CEF GS is also reported to be consistent with $|\pm 5/2\rangle$ state estimated by the temperature dependence of magnetic susceptibility \cite{nevidomskyy2009layered}.
Indeed, $|\pm 5/2\rangle$ state is plausible because the distribution of $4f$ hole extends toward neighboring heptagonal rings of boron atoms, maximizing the hybridization with boron rings.
However, no definitive experimental results including the estimation of the admixture with other $J_z$ components are reported.

\begin{figure*}
\includegraphics[bb=31 670 488 780]{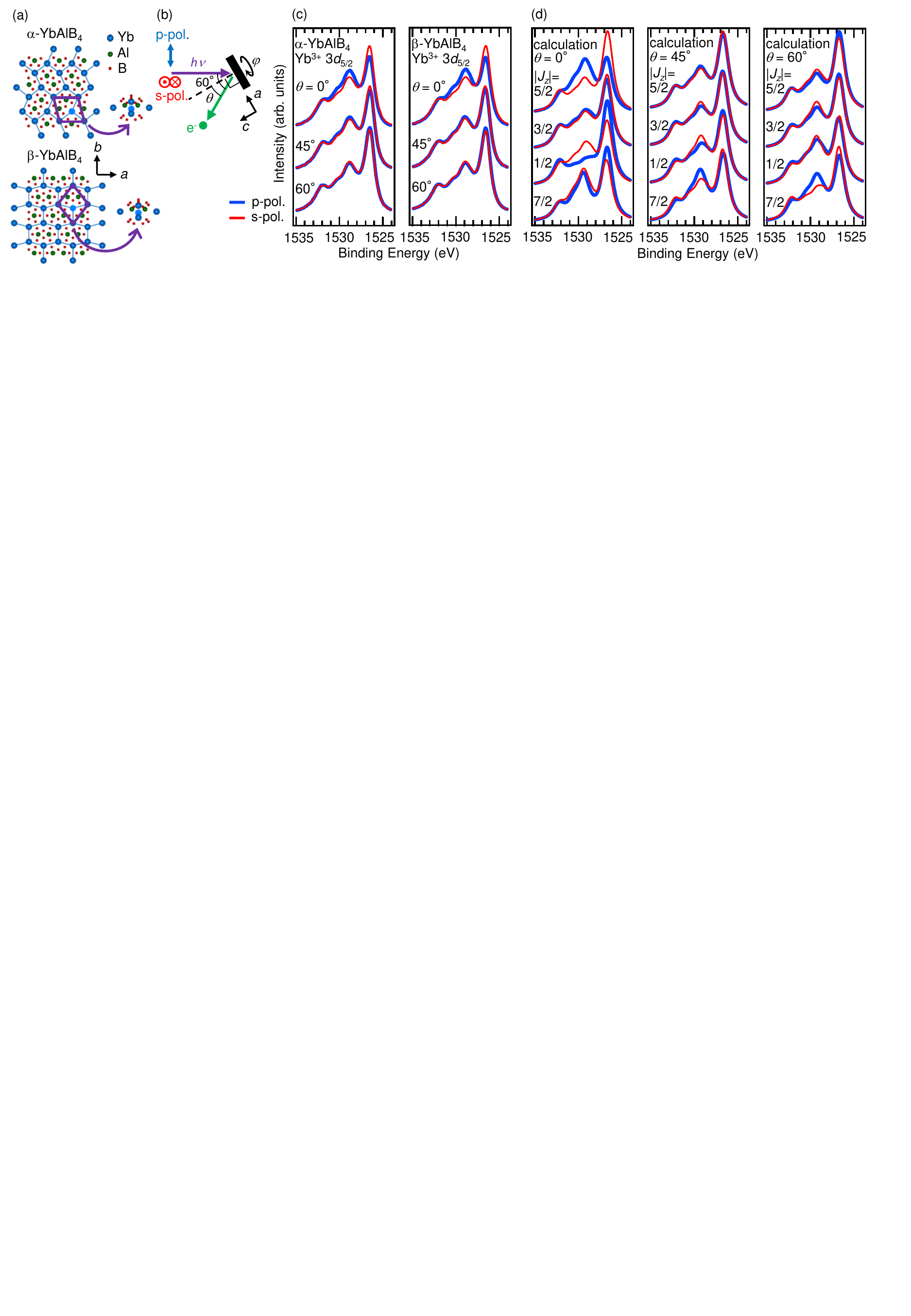}%
\vspace{18mm}
\caption{(a) Crystal structure of \abyb with a view along $c$-axis.
Yb-Al layer is sandwiched by heptagonal and pentagonal B layers, respectively.
Pictorial views inside the purple quadrangles which center the highlighted Yb atoms are shown at the right side.
The pictorial views orient to the acute angles of the purple quadrangles and the highlighted Yb atoms are at the front side.
(b) Schematic top view geometry of incident horizontally (blue arrow) and vertically (red arrow) polarized X-rays, sample and photoelectron analyzer.
The photoelectron analyzer is drawn as a photoelectron (e$^-$).
Polar angle $\theta$ is defined as an angle between the directions toward the photoelectron analyzer and $c$-axis of the sample, and $\theta = 60^{\circ}$ corresponds that the $c$-axis orients toward incident X-ray.
Azimuthal angle $\varphi$ is defined as the angle of the rotation along $c$-axis and $\varphi = 0^{\circ}$ corresponds to the geometry as drawn in Fig. 1(b).
(c) Polarization-dependent Yb$^{3+}$ $3d_{5/2}$ core-level HAXPES spectra of \abyb with the polar angle of $\theta = 0^{\circ}$, $45^{\circ}$ and $60^{\circ}$.
Red thin and blue thick lines correspond to s- and p-polarization.
The spectra are normalized by the spectral weight of Yb$^{3+}$ $3d_{5/2}$ after the subtraction of Shirley-type backgrounds \cite{Shirley72,Proctor82}.
(d) Ionic calculation of polarization-dependent Yb $3d_{5/2}$ core-level HAXPES spectra of pure $J_z$ states at each $\theta$.
\label{f1}}
\end{figure*}

Recently linear dichroism (LD) of angle-resolved core-level hard X-ray photoemission spectroscopy (HAXPES) has probed the CEF GSs of several Lanthanide based materials.
For example in Yb based materials, LDs of Yb $3d$ core-level multiplet structures determined the GSs of cubic system YbB$_{12}$ as $\Gamma _8$ and the GSs of tetragonal systems YbRh$_2$Si$_2$ and YbCu$_2$Si$_2$ as $|\pm 3/2\rangle$ and $-0.36|\pm 5/2\rangle + 0.93|\mp 3/2\rangle$, respectively \cite{Mori_LD,Kanai15}.
In this optical process, Yb $3d_{5/2}$ core-level photoemission reflects the charge distribution of Yb $4f$ hole through the interaction with created Yb $3d$ hole.
This method is also applicable to the orthorhombic system \byb.
In this letter, we show the experimental evidence of pure $|\pm 5/2\rangle$ CEF GSs in the orthorhombic compound \byb  probed by polar angle and azimuthal angle dependence of LDs of Yb 3$d$ HAXPES spectra.
In addition, we found the importance of explicit anisotropic hybridization between 4$f$ and conduction electrons to describe the CEF of \byb.
For comparison, we also measured stoichiometric compound \ayb which shows heavy Fermi liquid ground state \cite{matsumoto2011anisotropic}.

In \abyb, local symmetry at Yb site is $C_s$ and $C_{2v}$, respectively \cite{macaluso2007crystal}.
However, the regions surrounded by the four neighboring Yb sites resemble in both systems as shown in the purple quadrangles and their pictorial views in Fig. 1(a).
Therefore, we can approximate that the local symmetry at Yb site in \ayb is $C_{2v}$ and similar CEF GS is expected, which is turned out to be true later.
$J = 7/2$ states of Yb$^{3+}$ $4f$ electron in $C_{2v}$ symmetry will split into four doublet and $C_{2v}$ CEF is expressed by $B_2^0$, $B_2^2$, $B_4^0$, $B_4^2$, $B_4^4$, $B_6^0$, $B_6^2$, $B_6^4$ and $B_6^6$ in Stevens formalism \cite{Stevens}.
These parameters will produce the eight eigenfunctions expressed by using $J_z$ as $a_i|\pm 5/2\rangle+b_i|\mp 3/2\rangle+c_i|\pm 1/2\rangle+d_i|\mp 7/2\rangle (i = 1 \sim 4)$, where $a_i^2+b_i^2+c_i^2+d_i^2=1$.
For the simple comparison between measurement and calculation, we have performed ionic calculation including the full multiplet theory and the local CEF splitting using the XTLS 9.0 program \cite{3d_multiplet,XTLS}.
Atomic parameters such as the $4f$-$4f$ and $3d$-$4f$ Coulomb and exchange interactions (Slater integrals) and the spin-orbit couplings was obtained using Cowan's code based on the Hartree-Fock method \cite{Cowan81}.
The Slater integrals and spin-orbit couplings are reduced to 88\% and 98\% according to the references \cite{Yamaguchi09,Mori_LD,Kanai15}.
Single impurity Anderson model (SIAM) was also employed for the calculation which include explicit $c$-$f$ hybridization \cite{SIAM,SIAM_multiplet}.

We performed polarization-dependent Yb $3d$ core-level HAXPES with a MBS A1-HE hemispherical photoelectron spectrometer at BL19LXU of SPring-8 \cite{BL19LXU_instruments,BL19LXU_prl}.
Horizontally polarized X-ray radiation produced by 27 m long undulator was monochromated to be 7.9 keV by a Si(111) double-crystal and a Si(620) channel-cut crystal.
To switch the excitation light from horizontal to vertical polarization for LD, we used two diamond(100) single crystals as a phase retarder and vertically polarized component of the X-ray was 98\% \cite{azimuth}.
Since the direction of the photoelectron analyzer was in the horizontal plane with an angle about incident X-ray of $60^{\circ}$ as shown in Fig. 1(b), horizontally polarized and vertically polarized X-rays correspond to s- and p-polarization geometry, respectively.
The single crystals \abyb were grown by aluminum self-flux method \cite{macaluso2007crystal}.
Clean sample surfaces along $ab$-plane were obtained $in\: situ$ by cleaving under the pressure of $1 \times 10^{-7}$ Pa.
We have confirmed no impurities including oxygen and carbon by scanning core-level spectra.
The experimental geometry was controlled by using a two-axis manipulator which equips polar and azimuthal rotations \cite{azimuth}.
The energy resolution was set to $\sim$ 400 meV.
The sample temperature was set to 25 K, which is sufficiently lower than the first excited energy confirmed by negligible temperature dependence up to 60 K for \byb \cite{CEF_population} and by the much smaller entropy than $R$log2 estimated by specific heat \cite{matsumoto2011anisotropic}.

\begin{figure}
\includegraphics[bb=31 640 232 780]{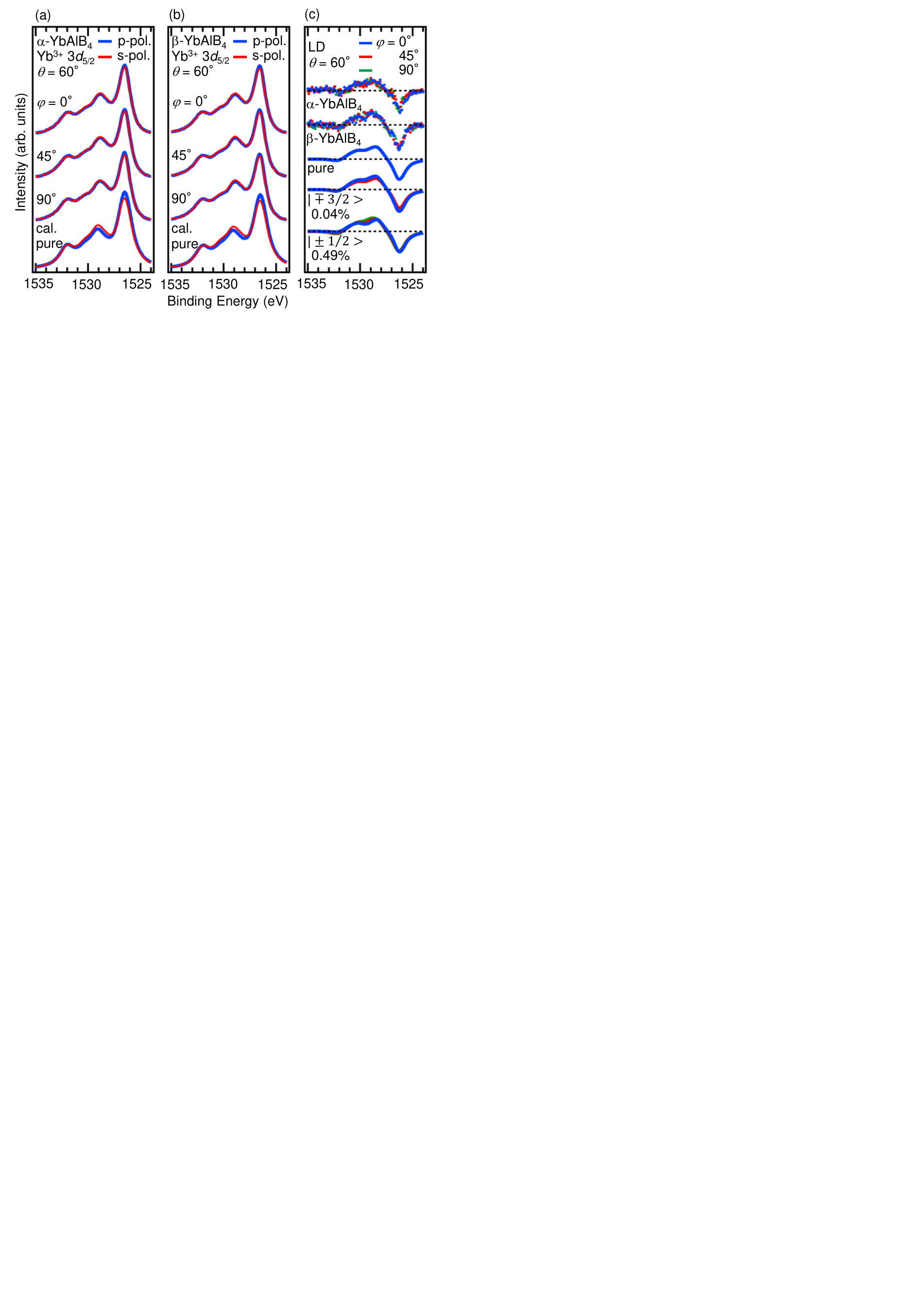}%
\vspace{16mm}
\caption{(a) and (b) The azimuthal angle dependence of Yb$^{3+}$ $3d_{5/2}$ core-level HAXPES spectra in \abyb at $\theta = 60^{\circ}$ and the ionic calculation of pure $|\pm 5/2\rangle$ state.
(c) The azimuthal angle dependence of LDs in \abyb (closed circle) and the ionic calculation for pure $|\pm 5/2\rangle$ state and $|\pm 5/2\rangle$ plus a small amount of $|\mp 3/2\rangle$ ($\sim 0.9998|\pm 5/2\rangle + 0.02|\mp 3/2\rangle$) or $|\pm 1/2\rangle$ ($\sim 0.9975|\pm 5/2\rangle + 0.07|\pm 1/2\rangle$) state (solid line).
LD is defined as the difference of normalized intensity between s- and p-polarization.
The amplitude of LDs by ionic calculation is divided by three to compare with experiments.
\label{f3}}
\end{figure}

Fig. 1(c) is the polarization-dependent Yb$^{3+}$ $3d_{5/2}$ core-level HAXPES spectra of \abyb with polar angle of $\theta = 0^{\circ}$, $45^{\circ}$ and $60^{\circ}$ as drawn in Fig. 1(b).
The multiplet structures peak at 1526, 1529 and 1532 eV with polarization and $\theta$ dependences which reflect the charge distribution of 4$f$ holes.
The polarization and $\theta$ dependences in \abyb are similar, suggesting almost the same CEF GS as expected by the neighboring atoms (Fig. 1(a)).
Peaks at 1526 eV with s-polarization geometry (s-pol.) are much higher, higher and lower than the case of p-polarization geometry (p-pol.) at $\theta = 0^{\circ}$, $45^{\circ}$ and $60^{\circ}$, respectively.
For the peaks at 1529 eV, these magnitude relations are reversed.
According to the ionic calculation shown in Fig.1 (d), all the magnitude relations are consistent with only $|\pm 5/2\rangle$ state.
However, the magnitude of the LDs which are defined as differences in the spectral intensities between s- and p- polarizations are about 3 times smaller than those of ionic calculation at $\theta = 0^{\circ}$ and $60^{\circ}$.
These comparisons suggest that the main component of the wave functions in \abyb is $|\pm 5/2\rangle$.
The possible reasons for the quantitative differences are the admixture with other $J_z$ components and/or $c$-$f$ hybridization because both effects will deform the charge distribution of $4f$ hole and will change LD.

Firstly, we focus on the mixing effect, namely $|\pm 5/2\rangle$ plus $|\mp 3/2\rangle$, $|\pm 1/2\rangle$ and/or $|\mp 7/2\rangle$.
Pure $|\pm 5/2\rangle$ state shows no rotational dependence in LDs along $c$-axis as shown in Fig. 2(c) because of the isotropic charge distribution around $c$-axis.
On the other hand, mixed states induce the azimuthal angle dependence as shown in CeCu$_2$Ge$_2$ \cite{CeCu2Ge2_Aratani} except the binary mixed state with $|\mp 7/2\rangle$.
However, it is unnatural if the wave function is $|\pm 5/2\rangle$ plus $|\mp 7/2\rangle$ because CEF Hamiltonian must have tetragonal symmetric terms of pure $|\pm 5/2\rangle$ state plus the six-fold symmetric term $B_6^6$ and other orthorhombic terms $B_2^2$, $B_4^2$ and $B_6^2$ must be zero.
Therefore we can distinguish if the CEF GS is pure $|\pm 5/2\rangle$ state or mixed state by measuring the azimuthal angle dependence of LDs.
Fig. 2(a), (b) and (c) show the Yb$^{3+}$ $3d_{5/2}$ core-level HAXPES spectra with different azimuthal angles $\varphi$ and with the fixed polar angle of $\theta = 60^{\circ}$ in \ayb, \byb and their LDs, respectively.
For both \abyb, LDs show azimuthal angle dependence within the experimental noise, indicating the pure $|\pm 5/2\rangle$ state.
To estimate the error bars of the wave function, we performed ionic calculation for the binary mixed states as shown in Fig. 2(c).
In case of $|\pm 5/2\rangle$ plus (minus) $|\mp 3/2\rangle$ state, four fold symmetric azimuthal angle dependence appears and LDs at $\varphi = 0^{\circ}$ and $90^{\circ}$ overlap and LD at $\varphi = 45^{\circ}$ is smaller (larger) than others.
The azimuthal angle dependence is very sensitive and experimental statistics suggests that the maximum possible component of $|\mp 3/2\rangle$ is only 0.04\%, indicating that LD by HAXPES is an extremely sensitive probe of electronic state.
In case of $|\pm 5/2\rangle$ plus (minus) $|\pm 1/2\rangle$ state, two fold symmetric azimuthal angle dependence appears and LD at 1529 eV increases (decreases) as $\varphi$ changes from $0^{\circ}$ to $90^{\circ}$.
The azimuthal angle dependence is less sensitive and experimental statistics suggests that the maximum possible component of $|\pm 1/2\rangle$ is 0.49\%.
If three or four components are mixed, $|\mp 7/2\rangle$ can have finite contribution, but the contribution will be much smaller than that of $|\mp 3/2\rangle$ and/or $|\pm 1/2\rangle$ states, resulting in the error bar much less than that of $|\mp 3/2\rangle$ or $|\mp 1/2\rangle$.
As these tiny error bars cannot explain the huge difference of LDs between the experiment and calculation by ionic model, we can conclude that the LDs in \abyb are modified by $c$-$f$ hybridization.
Note that the $c$-$f$ hybridization must be $\varphi$ independent because the anisotropy in $c$-$f$ hybridization will yield anisotropic LDs.

\begin{figure}
\includegraphics[bb=31 675 232 780]{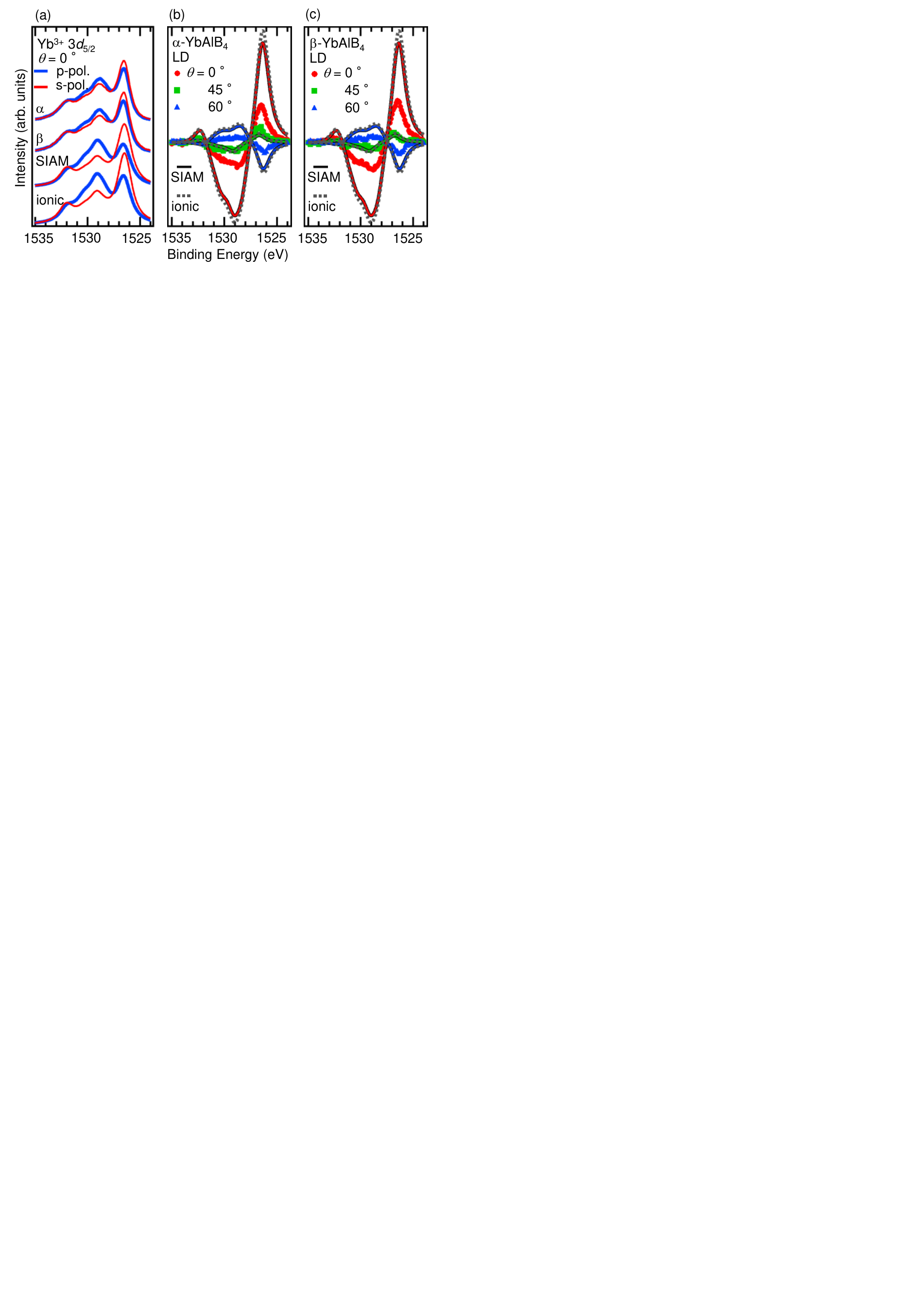}%
\vspace{18mm}
\caption{(a) Polarization-dependent Yb$^{3+}$ $3d_{5/2}$ core-level HAXPES spectra of \abyb at $\theta = \varphi = 0^{\circ}$, and the calculation of the spectra for $|\pm 5/2\rangle$ state based on SIAM including isotropic $c$-$f$ hybridization.
(b) and (c) The experimental LDs (closed circle, square and triangle) of \abyb and the calculation for $|\pm 5/2\rangle$ state based on SIAM (solid line) and ionic model (broken line).
Red, green and blue lines on the black lines correspond to the LDs by SIAM at $\theta = 0^{\circ}, 45^{\circ}$ and $60^{\circ}$.
Same parameters are used in the calculation shown in (b) and (c).
\label{f2}}
\end{figure}

As \abyb are strongly valence fluctuating systems with the Yb valences of 2.73 and 2.75 \cite{ybal-valency}, large $c$-$f$ hybridization is expected to modify LD, which was not apparently yielded in YbB$_{12}$, YbRh$_2$Si$_2$ and YbCu$_2$Si$_2$ \cite{Mori_LD,Kanai15}.
Next, we compare with the calculation based on SIAM for $|\pm5/2\rangle$ state in the mixed valence state between Yb$^{3+}$ (4$f^{13}$) and Yb$^{2+}$ (4$f^{14}\underline{L}$).
Here, $\underline{L}$ denotes the conduction band with one hole.
For simplicity, we employed isotropic $c$-$f$ hybridization.
Fig. 3(a) shows the Yb$^{3+}$ $3d_{5/2}$ core-level HAXPES spectra of \ayb, \byb and calculation by SIAM and ionic model at $\theta = \varphi = 0^{\circ}$.
The parameters of the Coulomb interaction between the $4f$ electrons and $3d$ core hole $U_{fc}$ and the effective $4f$ binding energy $\Delta _f$ and isotropic $c$-$f$ hybridization strength $V_{\rm{eff}}$ were set to be 10.0, 0.5 and 1.0 eV.
The peaks of multiplet by SIAM is slightly wider than those of ionic model and the multiplet structure will drastically change if the $V_{\rm{eff}}$ is larger than 2.0 eV (not shown).
Figure 3(b) and (c) show experimental LDs of \abyb and the calculated LDs by SIAM and ionic model.
For all geometry, calculations by SIAM show slightly smaller LDs than ionic model.
At $\theta = 0^{\circ}$ and $60^{\circ}$, LDs of calculation are still much larger than experimental results.
On the other hand, at $\theta = 45^{\circ}$, LD of calculation is smaller than that of experiment.
These comparisons indicate the importance of anisotropic $c$-$f$ hybridization to understand the CEF in \abyb which easily deforms the charge distribution as well as the LDs like oxides.
Furthermore we can see the different amplitude of LDs between \abyb: LDs of \ayb are smaller at $\theta = 0^{\circ}$ and $60^{\circ}$ and larger at $\theta = 45^{\circ}$, indicating the different anisotropy in $c$-$f$ hybridization which is integrated into CEF.

In summary, CEF GSs of \abyb are pure $|\pm 5/2\rangle$ states with the polar angle $\theta$ dependent and azimuthal angle $\varphi$ independent $c$-$f$ hybridizations.
These results remind us the topological non-trivial vortex metallic state originating from nodal hybridization in momentum space, which is predicted in \byb \cite{ramires2012beta}.
Experimental support in this nodal hybridization has been reported that laser ARPES measurement for \byb probed two anticrossing of 4$f$ CEF levels with dispersive conduction band which momentum dependence is consistent with nodal hybridization model \cite{laserARPES_YbAlB4}.
On the other hand, for \ayb, no clear Kondo hybridization band like \byb was observed by laser ARPES.
In our measurement of LD for \ayb, we observed clear $c$-$f$ hybridization in Yb 4$f$ $|\pm 5/2\rangle$ state possibly because Yb $3d_{5/2}$ core-level HAXPES reflects local feature compared with valence band measured by ARPES.
Furthermore, we discovered that the LD at each $\theta$ in \byb has a slightly smaller deviation from calculation than the LD in \ayb as shown in Fig. 3(b) and (c), indicating that \byb has the slightly smaller anisotropy of $c$-$f$ hybridization.
This small difference may correspond to the relation between the nodal and non-nodal hybridization as predicted in the previous theoretical study \cite{ramires2012beta}.
As we have shown the existence of anisotropy in $c$-$f$ hybridization, the detailed angle resolved core level LDs by HAXPES can be a new probe of anisotropic $c$-$f$ hybridization in real space.

\begin{acknowledgments}
We thank S. Fujioka, H. Aratani, T. Hattori, H. Yomosa, S. Takano, T. Kashiuchi, K. Nakagawa, K. Sakamoto and Y. Kobayashi for support in experiments. 
We also thank H. Kobayashi for useful discussions. 
Sample preparation was carried out under the Visiting Researcher's Program of the Institute for Solid State Physics, the University of Tokyo.
HAXPES experiments were performed at BL19LXU in SPring-8 with the approval of RIKEN (proposal nos. 20160034, 20160066, 20170043, 20170081, 20180026 and 20180076).
This work is supported by by a Grant-in-Aid for Scientific Research (16H04014, 16H04015, 18K03512), and a Grant-in-Aid for Innovative Areas (16H01074, 18H04317) from MEXT and JSPS, Japan.
This work is partially supported by CREST (JPMJCR15Q5,JPMJCR18T3), Japan Science and Technology Agency, by Grants-in-Aid for Scientific Research (16H02209), and by Grants-in-Aids for Scientific Research on Innovative Areas (15H05882, 15H05883) from MEXT. 
Y.K. was supported by the JSPS Research Fellowships for Young Scientists. 
\end{acknowledgments}

\bibliography{YbAlB4_LD.bib}

\end{document}